\documentclass[a4paper]{article}
\usepackage[pdftex]{graphicx}
\pdfoutput=1
\usepackage{geometry}
\newcommand{\be}{\begin{equation}}
\newcommand{\ee}{\end{equation}}
\newcommand{\bd}{\begin{displaymath}}
\newcommand{\ed}{\end{displaymath}}
\newcommand{\bea}{\begin{eqnarray}}
\newcommand{\eea}{\end{eqnarray}}
\newcommand{\bi}{\begin{description}}
\newcommand{\ei}{\end{description}}
\newcommand{\bq}{\begin{quote}}
\newcommand{\eq}{\end{quote}}

\begin{document}
\bibliographystyle{unsrt}
\author{Alexander~Unzicker\\
        Pestalozzi-Gymnasium  M\"unchen\\[0.6ex]\\
{\small{\bf e-mail:}  alexander.unzicker@lrz.uni-muenchen.de}}

\author{
Alexander~Unzicker\\
         {\small Pestalozzi-Gymnasium M\"unchen, Germany}\\
         {\small aunzicker@web.de}}
\title{Immanuel Kant on Supersymmetry: \\ A Practical Evaluation}
\maketitle

\begin{abstract}
A short review of the motivations for supersymmetry in astrophysics
and particle physics is given. Despite the amount of theoretical research
conducted in the past decades, no observational evidence for supersymmetry
has yet been found. While a large part of the community is expecting
supersymmetry to be discovered in the Large Hadron Collider (LHC),
some of the basic arguments in favor are disputed here. Since it is not excluded
that the author's view may be biased by his research, he proposes
a bet on the discovery of supersymmetric particles: 
According to the philosopher Immanuel Kant, the bet marks the difference
between persuasion and conviction.

\end{abstract}

\paragraph{Supersymmetric particles as dark datter candidates.}

The Swiss astronomer Fritz Zwicky observed as early as 1933 that the 
velocities of individual galaxies in clusters were much higher than
expected. The gravitational potential of the visible matter would have been insufficient
to keep them within the cluster - the first hint towards the existence of
`dark matter'. In the meantime, numerous observations indicate that
only a small part of matter in the universe is visible. Flat rotation curves of galaxies, 
overly hot gas in clusters detected by X-rays, lensing results, and the
homogeneity of the cosmic microwave background, which cannot account
for the observed structure formation. In the latter case, even an extra 
assumption about fluctuations of dark matter had to be invoked.

From a general perspective, many results are contradictory, and
still do not fit into the common picture \cite{Unz:07}. Even the
celebrated discovery of dark energy is sometimes considered as a post-hoc-fix
involving a new free parameter \cite{Agu:01a}.
In particular, the issue of small accelerations has received some attention:
can we be sure about the validity of the law of gravitation for accelerations
in the regime $10^{-10} m s^{-2} \approx \frac{c}{T_u}$ ($T_u$ being the age
of the universe)~?
The Pioneer anomaly \cite{And:01, Unz:07, Lam:06, Tur:10} hints in that
direction, as well as velocity dispersions in globular clusters \cite{Sca:07}.
Many experts on galaxy dynamics claim that their observations cannot be 
explained by particles whatsoever \cite{Sal:07, Eva:01, Sel:00}.
The greatest obstacle for dark matter theories is the sucessful phenomenological description
of rotation curves by MOND \cite{Mil:84, Agu:01a, San:02}. Although this theory is
not convincing at all, the appearance of the enigmatic acceleration 
of $10^{-10} m s^{-2}$ allows only two conclusions: either more than 1000 
spiral galaxies agreed on fooling today's astronomers or gravity is not yet
understood on a fundamental level.
To summarize, the astrophysical evidence for the existence of supersymmetric
particles seems to be much less than commonly assumed.
A closer look at the details reveals that such an explanation is somewhat naive. 
 
\paragraph{Tired of the standard model.}

Some parameters of the standard model of particle 
physics vary with energy; this phenomenon 
is called `running constants'. Though the energy
range of the observations, considering logarithmic scales,
is not large, the trend of the coupling constants of the basic
interactions can be extrapolated. \cite{Ama:91} Unfortunately,
the three straight lines do not cross in one point. 
According to supersymmetric models, this crossover
can be achieved by a particle in the range of $\sim 1 \ TeV$.
In that case, one may hope for a `unification' of 
the interactions at high energies. Though this
paradigma requires further assumtions like the
`desert hypothesis' (nothing interesting happens in the 
energy range in between), this type of extrapolation
seems ambitious, if not ridiculous:
11 orders of magnitude, and a justification 
by error analysis is still missing.

\paragraph{Theoretical reasons.}

The standard model claims a unification of the electromagnetic,
the weak and the strong force by means of the symmetry
group $U(1) \times SU(2) \times SU(3)$. However, 
this is merely a framework of analogies,
in which many elements are still missing. Sheldon Glashow,
Nobel prize winner for the standard model, said \cite{Dav}:
\begin{quote}
This theory is an ad hoc construction, we had to insert a couple of things
which 
are still mysterious. For instance, why do
 the particle masses have the values we observe~?'
\end{quote} 

More drastically, this was emphazised by Richard Feynman \cite{Gle}:
\begin{quote}
`Three theories. Strong interactions, weak 
interactions, and the electromagnetic. . . . The theories are 
linked because they seem to have similar characteristics. . . .
 Where does it go together? Only if you add some stuff we don’t know.
 There isn’t any theory today that has $SU(3) \times SU(2) \times U(1)$ — 
whatever the hell it is — that we know is right, that has any 
experimental check... Now, these guys are all trying to put all
 this together. They're {\em trying\/} to. But they havn't. Okay?'

\end{quote} 

Given this, it seems that theoreticians do not sufficiently 
reflect whether the symmetry group approach can satisfactory
be filled with physics. Howard Georgi, who dealed with 
great unified theories and therefore not a suspect 
of fundamentalists's criticism, said:

\begin{quote}
`Symmetry is a tool for finding the underlying dynamics, 
which 
subsequently has to explain the success (or failure)
of the symmetry 
arguments. Group theory is a useful method, 
but does not substitute physics.'
\end{quote} 

Such reflections seem to die out, and despite Wolfgang Pauli's 
harsh expression  `group pest',
group theory has become the dominant language of
particle physics. 

Since the group $SU(5)$ beautifully embraces the 
glue-construct $U(1) \times SU(2) \times SU(3)$, supersymmetry
had been considered as a natural extension of the 
standard model. 
According to that hypothesis, to every boson (integer spin) there
should exist a partner fermion (half-integer spin), and vice versa.
The partners are named with a s-prefix or suffix -ino.
But how do a `selectron' and a `sproton' form a `satom',
since the hydrogen atom is a boson? It seems somehow naive
to believe that nature just doubles its properties,
though a prominent precursor for such a process exists:
Paul Dirac had predicted the positron and the antiproton
before their discovery. However, these partners have 
- besides their opposite charge - otherwise identical properties,
in particular masses, while the necessarily different
masses of superpartners are usually explained by 
a `broken symmetry'. But how beautiful is a symmetry 
that breaks as soon as it meets the facts~?

\paragraph{Open issues.}

The problem that the standard model of particle physics
is unable to calculate masses, is not ameliorated
by supersymmetry, even worse: while for the Higgs boson,
the desired but missing ingredient of the standard model,
at least an energy range is predicted. Supersymetry, on
the other hand, does not predict masses at all. 
If a particle is discovered,
it is fine, if not, the theory will survive at higher
energies. According to the fairplay rules of scientific
methodology given by Karl Popper, this is dangerously
close to non-falsifiability: the theory behaves 
like someone who likes winning, but not loosing.

Richard Feynman stresses the same point in his sarcastic
fashion \cite{FeyQED}:

\begin{quote}
`Somebody makes up a theory: The proton is unstable. 
They make a 
calculation and find that there would be no protons in the
 universe 
any more! So the fiddle around with their numbers, putting 
a higher mass
into the new particle, and after much effort they predict
 that 
the proton will decay at a rate slightly less than the last 
measured 
rate the proton has shown not to decay at.'
\end{quote} 

Last but not least, postulating about 100 freely adjustable
constants, supersymmetry reminds us from an erosion of
scientific theories described by the philosopher Thomas
Kuhn: such an increase of unexplained numbers was the
characteristic attribute of the geocentric view of 
the world which led to the well-known
epicyles of the Ptolemaic theory.

\paragraph{Kant, conviction and the bet.}

To summarize, we find that a mixture of physically motivated aspirations
and purely mathematical reasoning led to $> 40000$ publications
in the past decades, according to SPIRES \cite{Woi:06}.
Given that up to now there is no experimental 
verification of any prediction made, this could 
lead to sociological considerations, but given that such
a strong conviction of the existence of supersymmetry lives on,

one should try to measure it.
The German philosopher Immanuel Kant wrote in 
his most famous work, {\em The Critique of Pure Reason\/}:

\begin{quote}
The usual test,
 whether that which any one maintains is merely 
his persuasion, or his 
subjective conviction at least, that 
is, his firm belief, is a bet. It 
frequently happens that a 
man delivers his opinions with so much boldness and
 assurance, that he appears to be under no apprehension as to the 
possibility of his being in error. The offer of a bet startles
 him, and makes him pause. Sometimes it turns out that his 
persuasion may be valued at a ducat, but not at 
ten. For he does not hesitate, perhaps, to venture a ducat, but if it 
is proposed to stake ten, he immediately becomes aware of 
the possibility of his being mistaken - a possibility which 
has hitherto escaped his observation. If we imagine to 
ourselves that we have to stake the happiness of our whole 
life on the truth of any proposition, our judgement drops 
its air of triumph, we take the alarm, and discover the 
actual strength of our belief. Thus pragmatical 
belief has degrees, varying in proportion to the interests at stake.

\end{quote}

Nicholos Wethington, in an article in {\em Universe Today\/} \cite{Wet:10},
commented: 

\begin{quote}
If you haven't had the fortune (as have I) of four years studying philosophy, 
this passage from Kant can be neatly summed up with the old adage, `Put your money where your mouth is.'
\end{quote}

A modern possibility how to realize Kant's proposal are prediction markets
like Intrade.com. Contracts on the discovery of supersymmetric particles
can be traded there like stock options, which allows bet both on the
discovery and the non-discovery. Since the first results from the LHC are
expected soon, it is time for the `pragmatical belief' to be evaluated.

\paragraph{How to bet on a supersymmetric particle at a prediction market.}\footnote{A
corresponding description for the Higgs boson is given at the site {\em www.Bet-On-The-Higgs.com\/}}.

A prediction market like Intrade is based on the trade of contracts
on given events. The idea is that the actual price
is a measure of the probability that the event will happen.
If you belive for instance that a supersymmetric particle will be
discovered until the end of 2011, you may buy the
contract SUSY.PARTICLE.DEC11. It is noteworthy that you may 
even {\em sell} contracts that you do not hold, in case you do not
believe the event will happen.

After opening an account and uploading a deposit 
one may start betting immediately. 
Contracts have a nominal value of \$10 corresponding to 100 points.
You may buy one of the contracts SUSY.PARTICLE.DEC10
or even SUSY.PARTICLE.DEC13 if you believe the discovery 
 will occur within the period \footnote{The relevant date is the 
date of publication.} or sell, if you don't believe.
As in a stock market, there is a bid and an ask 
price.\footnote{{\em http://en.wikipedia.org/wiki/Ask\_price\/}}
If the event happens, the contract achieves the value 100, if not, zero.
This relates the current rate directly to the percent probability 
that the event will happen.
All other technical details how to bet there
can be found in \cite{Unz:09} which describes a corresponding bet
on the Higgs boson. 

It is suggested that the use of prediction markets in science can 
enhance the evaluation of research.

\end{document}